\documentclass[floatfix,prd,superscriptaddress,nofootinbib,amsmath,amssymb,aps,twocolumn]{revtex4-2}
\usepackage{graphicx}
\usepackage{dcolumn}
\usepackage{bm}
\usepackage{mathrsfs}
\usepackage{xcolor,graphicx}
\usepackage{dcolumn}
\usepackage{bm}
\usepackage{enumerate}
\usepackage{feynmf}
\usepackage{subfigure}
\usepackage{todonotes}
\usepackage{multirow}
\usepackage{hyperref}
\usepackage{xspace}
\usepackage{float}
\usepackage{ulem}
\newcommand{\mg}{\texttt{MadGraph5\_aMC@NLO}\xspace}
\newcommand{\pythia}{\texttt{PYTHIA~8}\xspace}
\newcommand{\delphes}{\texttt{DELPHES~3}\xspace}
\newcommand{\eec}{$e^+e^-$ collider\xspace}

\newcommand{\pt}{p_\text{T}}
\newcommand{\gev}{\operatorname{GeV}}

\newcommand{\fb}{\operatorname{fb}}

\newcommand{\br}{\operatorname{Br}}

\allowdisplaybreaks[4]

\begin{document}

\title{Exposing new scalars hiding behind the Higgs boson}

\author{Qing-Hong Cao}
\email{qinghongcao@pku.edu.cn}
\affiliation{Department of Physics and State Key Laboratory
of Nuclear Physics and Technology, Peking University, Beijing 100871, China}
\affiliation{Center for High Energy Physics, Peking University, Beijing 100871, China}

\author{Kun Cheng}
\email{chengkun@pku.edu.cn}
\affiliation{Department of Physics and State Key Laboratory
of Nuclear Physics and Technology, Peking University, Beijing 100871, China}

\author{Yandong Liu}
\email{ydliu@bnu.edu.cn}
\affiliation{Key Laboratory of Beam Technology of Ministry of Education, College of Nuclear Science and Technology, Beijing Normal University, Beijing 100875, China}
\affiliation{Beijing Radiation Center, Beijing 100875, China}

\author{Xin-Kai Wen}
\email{xinkaiwen@pku.edu.cn}
\affiliation{Department of Physics and State Key Laboratory
of Nuclear Physics and Technology, Peking University, Beijing 100871, China}

\author{Changlong Xu}
\email{changlongxu@pku.edu.cn}
\affiliation{Department of Physics and State Key Laboratory
of Nuclear Physics and Technology, Peking University, Beijing 100871, China}

\author{Hao Zhang}
\email{zhanghao@ihep.ac.cn}
\affiliation{Theoretical Physics Division, Institute of High Energy Physics, Chinese Academy of Sciences, Beijing 100049, China}
\affiliation{School of Physics, University of Chinese Academy of Sciences, Beijing 100049, China}
\affiliation{Center for High Energy Physics, Peking University, Beijing 100871, China}

\begin{abstract}
It is possible that there is another scalar hiding behind the known 125 GeV Higgs boson. If the hidden scalar exhibits a CP property different from the Higgs boson, it can be exposed in the di-Higgs production at the high-luminosity large hadron collider and future colliders. 
\end{abstract}

\maketitle

\section{Introduction}

The Higgs boson in the Standard Model (SM) is responsible for the electroweak symmetry breaking (EWSB). It is invariant under the CP transformation, where C stands for Charge conjugation, and P denotes Parity reflection. Although only a neutral CP-even Higgs boson is present in the SM, more than one Higgs boson may exist in many new physics (NP) models. A CP-odd scalar is a definite signal of new physics and helps us understand the EWSB mechanism. Many efforts have been made in searching for new CP-odd or CP-mixture scalar resonances, but no new heavy scalars have been found, and we end up with only one scalar around 125 GeV, known as the SM-like Higgs boson.  
Since its discovery, various properties of the Higgs particle have been accurately measured. In numerous experimental measurements, it is often assumed that there is only one Higgs boson.
{\it But is there another new scalar hiding behind the Higgs boson?} For example, the new scalar has a mass degenerate with or very close to that of the Higgs boson, but the Higgs boson covers it up in single Higgs-boson productions such that one cannot recognize it directly. If so, how to detect it? This work shows that the hidden scalar can be exposed in di-Higgs production as long as it has a CP property different from the Higgs boson.

The key is the $Zhh$ interaction that is absent in the SM. 
The general $Zh_1h_2$ interaction reads as 
\begin{align}
	\mathcal{L} \supset i g_{Zh_1h_2}\left(h_1 \partial^{\mu} h_2-h_2 \partial^{\mu} h_1\right)Z_{\mu},
\end{align}
where $g_{Zh_1h_2}$ is the effective coupling. Without loss of generality, we denote $h_1$ and $h_2$ as the known Higgs boson and the hidden scalar, respectively, and 
\begin{equation}
    m_{h_1}=m_{h_2}=125~\gev.    
\end{equation}
The $Zh_1h_2$ coupling is nonzero only when $h_1$ and $h_2$ are not identical, and
in addition, the two scalars must exhibit different CP properties. There are four possibilities~\footnote{The possibility that the Higgs boson is a CP-odd scalar has been excluded at $99\%$ confidence level~\cite{CMS:2013fjq,ATLAS:2013xga}.}: 
\begin{itemize}
\item a CP-even $h_1$ and a CP-odd $h_2$, 
\item a CP-even $h_1$ and a CP-mixing $h_2$, 
\item a CP-mixing $h_1$ and a CP-odd $h_2$, 
\item both $h_1$ and $h_2$ are CP-mixing but with different CP angles.
\end{itemize}
Therefore, observing a $Zh_1h_2$ coupling provides evidence of a new scalar degenerate with the known Higgs boson, and more importantly, there must be a CP-odd scalar component.

The best way to measure the $Zh_1h_2$ coupling at colliders is through the electroweak di-Higgs boson production, namely $f\bar{f}\to Z^\ast\to h_1h_2$, where $f$ stands for quarks or leptons. 
Given the fermion $f$ coupling to $Z$ gauge boson of $\bar{f}(ig_V^{f}\gamma_\mu+ig_A^{f}\gamma_\mu\gamma_5)f Z^\mu$, where $g_V^{f}$ denotes the vector-like coupling and $g_A^{f}$ stands for axial-vector-like one, the cross-section with the collision energy of $\sqrt{s}$ is
\begin{align}\label{eq:dsdcosffZhA}
    {\hat\sigma}_{f\bar{f}}(s)=\frac{g_{Zh_1h_2}^2[(g_A^{f})^2+(g_V^{f})^2](s-4m_h^2)^{3/2}}{48\pi s^{1/2} (s-m_Z^2)^2},
\end{align}
and the couplings of SM fermions $f$ to $Z$ gauge boson read as 
\begin{align}
g_V^f=\frac{g}{\cos\theta_W}\biggl(\frac{T_3^f}{2}-Q_f\sin^2\theta_W\biggr),\quad g_A^f=-\frac{g}{\cos\theta_W}\frac{T_3^f}{2}, \nonumber
\end{align}
where $g$ is the weak gauge coupling, $\theta_W$ is the Weinberg angle, $Q_f$ denotes the electric charge of the fermion $f$, $T_3^f$ is $1/2$ for up-type quarks (neutrinos) and $-1/2$ for down-type quarks (charged leptons). 
At the proton-proton ($pp$) collider the cross-section of $pp\to Z^\ast\to h_1h_2$ reads as 
\begin{align}
    \sigma_{pp}(S)=\sum_{q} \int &d x_1 d x_2 \biggl\{f_{q{/p}}\left(x_1, \mu_F\right) f_{\bar{q}{/p}}\left(x_2, \mu_F\right)\nonumber\\ &\times\hat{\sigma}_{q\bar{q}}\left({x_1x_2S}\right)
    +(x_1\leftrightarrow x_2)\biggr\},
\end{align}
where 
$f_{a/p}\left(x, \mu_F\right)$ is the parton distribution function (PDF) of finding a parton $a$ inside a proton with a momentum fraction of $x$, and $\sqrt{S}$ is the energy in the center of momentum (COM) frame of the proton-proton collision.

The paper is organized as follows. In Sec.~\ref{sec:lhc}, we examine the discovery potential of the coupling $g_{Zh_1h_2}$ through the di-Higgs boson production. Furthermore, we explore the measurement of the coupling $g_{Zh_1h_2}$ at the future hadron colliders in Sec.~\ref{sec:FCC-hh} and at the lepton colliders in Sec.~\ref{sec:cepc}, respectively. Finally, we conclude in Sec.~\ref{sec:conclusion}.

\section{Discovery potential at the LHC}
\label{sec:lhc}

We begin with 13 TeV LHC. The cross-section of the di-Higgs production through the process of $pp \to Z^*\to h_1 h_2$ is $$\sigma_{Zh_1h_2}=890.0\times g_{Zh_1h_2}^2~\text{fb},$$
which includes the NLO QCD correction calculated by $\mathtt{MadGraph5\_aMC@NLO}$~\cite{madgraph5} with $\mathtt{CT18NLO}$ PDF~\cite{Hou:2019efy}.
In the SM, the di-Higgs boson production is dominated by the gluon-gluon fusion (ggF) process of $gg\to hh$ at 13 TeV LHC. The experimental collaborations have conducted extensive research on the channel with various decay modes of the Higgs bosons~\cite{ATLAS:2022jtk,ATLAS:2022fpx,Mukherjee:2022mua,CMS:2022kdx,CMS:2022hgz}. Currently, an upper limit of $3.4\sigma_{\rm SM}^{hh}$ at the $95\%$ confidence level (C.L.) has been achieved on the di-Higgs boson production after the combination of each Higgs boson decay mode with full LHC run-II data~\cite{CMS:2022dwd}. The limit according to the $b\bar{b}\gamma\gamma$ decay mode is $8.4\sigma_{\rm SM}^{hh}$, where $\sigma_{\rm SM}^{hh}{=32.76^{+1.95}_{-6.83}}\fb$ is the SM prediction of di-Higgs boson production cross-section at 13 TeV LHC, including the next-to-leading order~(NLO) QCD corrections~\cite{Grazzini:2018bsd,Baglio:2020wgt,Dreyer:2018qbw}.

We focus on the $b\bar{b}\gamma\gamma$ decay mode due to its clear signal at hadron colliders. The number of signal events, including contribution from the electroweak process, reads as 
\begin{align}
    N_{\rm obs}
    =&N^{hh}_{\rm SM}\left(1 + \frac{\sigma_{Zh_1h_2}}{\sigma^{hh}_{\rm SM}} \kappa^{h_1h_2}_{b\bar{b}\gamma\gamma}\kappa_{\rm select}\right),
\end{align}
where $\kappa^{h_1h_2}_{b\bar{b}\gamma\gamma}$ is the $b\bar{b}\gamma\gamma$ decayed fraction of $h_1h_2$ pair with respect to that of the SM $hh$ pair, i.e.
\begin{equation}\label{eq:kappadecay}
    \kappa^{h_1h_2}_{b\bar{b}\gamma\gamma}=\frac{\br_{h_1\to b\bar{b}} \br_{h_2\to \gamma\gamma}+\br_{h_2\to b\bar{b}} \br_{h_1\to \gamma\gamma}}{2 \br_{h_{\rm SM}\to b\bar{b}} \br_{h_{\rm SM}\to \gamma\gamma}},
\end{equation}
and $\kappa_{\rm select}$ is the ratio of the cut efficiencies between the $h_1h_2$ pair production and the SM di-Higgs production process. In this work, we assume that the net cross-section of the processes of $gg\to h_{1/2}h_{1/2}$ is the same as $\sigma_{\mathrm{SM}}^{hh}$. Of course, the cross-sections of $gg\to h_{1/2}h_{1/2}$ are sensitive to the Yukawa interactions of the Higgs boson and hidden scalar to the SM quarks in specific new physics models; however, we focus on a model-independent study in this work and will present the study of various new physics models elsewhere~\cite{Cao:2022pre}. 

With the assumption of the same decay branching ratio with the SM Higgs boson and cut efficiency, namely $\kappa_{b\bar{b}\gamma\gamma}^{h_1h_2}=\kappa_{\rm select}=1$, it is found that the current constraint on the di-Higgs boson production rate can be converted on the $\sigma_{Zh_1h_2}$, e.g., $\sigma_{Zh_1h_2}<7.4 \sigma_{\rm SM}^{hh}$ from the $b\bar{b}\gamma\gamma$ decay mode, and the constraint can be transferred to a bound on the effective coupling $g_{Zh_1h_2}$ as
\begin{align}
    g_{Zh_1h_2}<0.53.
\end{align}
Combining all the Higgs boson decay modes yields a stricter bound  $g_{Zh_1h_2}<0.30$.

Next, we are devoted to the high-luminosity LHC (HL-LHC) with a collision energy of $\sqrt{S}$=14 TeV. The projection from the ATLAS collaboration yields 8.4 signal events and 47 background events based on the $b\bar{b}\gamma\gamma$ signal for the SM di-Higgs boson process at the HL-LHC. The significance is expected to be around $1.3\sigma$~\cite{ATL-PHYS-PUB-2014-019} and can be improved by benefiting from machine learning technique~\cite{ATLAS:2018dpp,Monti:2788631}. However, constructing a machine-learning algorithm is beyond the scope of this work. We still follow the traditional cut-based analysis.

We generate the leading-order parton-level signal events using \mg~\cite{madgraph5} with $\mathtt{CT18LO}$ PDF~\cite{Yan:2022pzl}, pass the events to \pythia~\cite{pythia} for parton showering and hadronization, and then simulate the detector effects with \delphes~\cite{delphes}. 
We choose the same selection criteria with Ref.~\cite{ATL-PHYS-PUB-2014-019}, and the preselection cuts are
\begin{align}
& 2\leq  n^j<6,\quad p_\text{T}^j>25~{\rm{GeV}},\quad |\eta^j|<2.5,\nonumber\\
& n^\gamma\geq 2,\quad p_\text{T}^\gamma>30~{\rm{GeV}},\quad |\eta^\gamma|<2.37,\nonumber\\
& \Delta R^{\gamma \gamma}>0.4,\quad \Delta R^{\gamma j}>0.4,\quad \Delta R^{j j}>0.4,
\end{align}
where $j$ denotes jet reconstructed using the anti-$k_{\mathrm T}$ algorithm with jet radius parameter $R = 0.4$, and 
\begin{align}
\Delta R^{mn} = \sqrt{(\eta^m-\eta^n)^2+(\phi^m-\phi^n)^2}.
\end{align}
As in the signal events, there are two bottom-quarks from the Higgs boson decay, it is demanded the first two leading jets must be tagged as $b$-jets with a mean tagging efficiency of 70\%, and
\begin{align}
   & p_\text{T}^{b_1}>40~{\rm{GeV}},\quad p_\text{T}^{b_2}>25~{\rm{GeV}},\nonumber\\
   & 100~{\rm{GeV}}<m_{b_1b_2}<150~{\rm{GeV}},
\end{align}
where $b_1$ and $b_2$ denote the leading and the sub-leading $b$-tagged jet, respectively. Similarly, the two photons decaying from the other Higgs boson in the signal events and the invariant mass of the leading ($\gamma_1$) and the sub-leading photon ($\gamma_2$) must satisfy
\begin{align}
     123~{\rm{GeV}}<m_{\gamma_1\gamma_2}<128~{\rm{GeV}}.
\end{align}
In addition, the transverse momenta of the $b_1 b_2$-system and the diphoton system are both required to satisfy
\begin{align}
    p_\mathrm{T}^{\gamma_1\gamma_2},~p_\mathrm{T}^{b_1b_2}>110~\mathrm{GeV}
\end{align}
to suppress the QCD backgrounds.
The efficiency of the $pp\to Z^\ast \to h_1h_2$ process is almost the same as the ggF process, and the ratio is $\kappa_{\rm select}= 0.94$.
The NLO QCD corrected production cross-section is
$$\sigma_{Zh_1h_2}=992.1\times g_{Zh_1h_2}^2~\text{fb}$$
at the HL-LHC. So the observed $b\bar{b}\gamma\gamma$ event number from $h_1 h_2$ pair production is
\begin{equation}
185\times g_{Zh_1h_2}^2 \kappa^{h_1h_2}_{b\bar{b}\gamma\gamma},    
\end{equation}
where $\kappa^{h_1h_2}_{b\bar{b}\gamma\gamma}$ is defined in Eq.~\ref{eq:kappadecay} to describe the di-Higgs boson decay branching ratio to $b\bar{b}\gamma\gamma$ in contrast to SM value. The background event number is 47, the same as in~Ref.~\cite{ATL-PHYS-PUB-2014-019}.

\begin{figure}
	 \centering
	\includegraphics[scale=0.6]{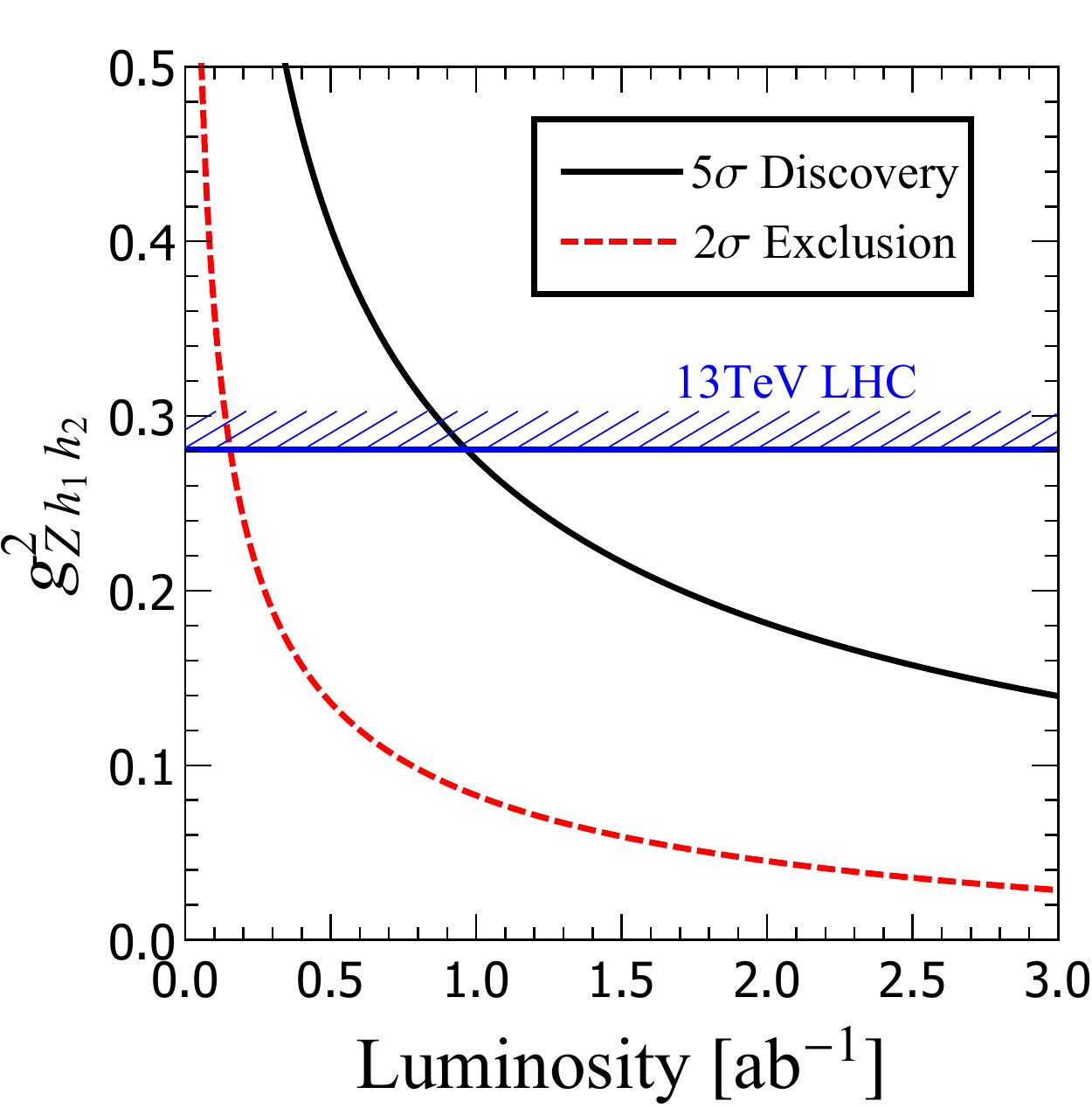}
    \caption{The contours of discovery potential of $pp\to Z^*\to h_1h_2$ events at the HL-LHC in the plane of $g_{Zh_1h_2}^2$ and the integrated luminosity, where the branch ratios of $h_{1,2}$ decays are chosen as the same as those of the SM-like Higgs boson. The region above the blue line is excluded by the LHC data. 
    }
	\label{fig:HLLHCgL}
\end{figure}

Equipped with signal and background event numbers, the significance of HL-LHC potential on the di-Higgs boson production could be estimated as
\begin{equation}
    z= \frac{n_s}{\sqrt{n_b}}=\left(1.23+27.0 g_{Z h_1h_2}^2  \kappa^{h_1h_2}_{b\bar{b}\gamma\gamma}\right)\sqrt{\frac{L}{3000~{\rm fb}^{-1}}} ,
\end{equation}
where $n_s$ and $n_b$ denote the signal and background event numbers, respectively.  Figure~\ref{fig:HLLHCgL} displays the significance contour in the plane of $g_{Zh_1h_2}^2$ and the integrated luminosity with the assumption of $\kappa_{b\bar b\gamma\gamma}^{h_1h_2}=1$. The black curve denotes the $5\sigma$ discovery potential. The horizontal blue line presents the constraint from 13~TeV LHC, and the region above the blue line is excluded. The current limit of $g_{Zh_1h_2}=0.53$ (single channel) and $g_{Zh_1h_2}=0.30$ (combined analysis) can be discovered with an integrated luminosity of $1~\rm{ab}^{-1}$ and $5.6~\rm{ab}^{-1}$ at $5\sigma$ C.L., respectively. The red dashed curve represents the exclusion limit at $2\sigma$ C.L. 
It shows that if no deviation was observed in the di-Higgs production, one could obtain a tighter constraint of $g_{Zh_1h_2}\leq 0.15$ when an integrated luminosity of $3~\rm{ab}^{-1}$ is achieved.

\section{Future hadron colliders}
\label{sec:FCC-hh}

In this section, we consider a future 100 TeV $pp$ collider, named FCC-hh,
which is supposed to measure the di-Higgs boson production precisely. 
The NLO cross-section of the process of $q\bar{q}\to Z^*\to h_1h_2$ is
\begin{equation}
\sigma_{Zh_1h_2}=12010.0\times g_{Zh_1h_2}^2~\text{fb}.
\end{equation}

The parameters for the collider simulation are adopted as proposed in Ref.~\cite{Contino:2016spe,Mangano:2016jyj}.
The jets are reconstructed using the anti-$k_{\mathrm T}$ algorithm with jet radius parameter $R = 0.5$ and the $b$-tagging efficiency is chosen to be 75\%.
It is required that there are at least two photons and two $b$-tagged jets in signal events, and
\begin{align}
    p_\text{T}^{b_1,\gamma_1}>60\gev,\quad p_\text{T}^{b_2,\gamma_2}>35\gev,\quad|\eta_{\gamma,b}|<4.5,
\end{align}
where $b_1 (\gamma_1)$ denotes the leading transverse momentum $b$-jet (photon) and $b_2 (\gamma_2)$ the sub-leading one. 
The invariant mass cuts are chosen to be
\begin{align}
 &100\gev<   m_{b_1b_2}<150\gev,\\
    &123\gev<   m_{\gamma_1\gamma_2}<127\gev,
\end{align}
as the bottom-quark pair and photon pair are either from the Higgs boson or the hidden scalar in signal events. For the same reason, we required that the separation $\Delta R$ of two $b$-jets and two photons is not too large, namely
\begin{align}
    \Delta R^{b_1b_2},~\Delta R^{\gamma_1\gamma_2}<3.5,
\end{align}
and the transverse momentum of the $b_1b_2$-system and the diphoton system is required to be larger than 100 GeV, namely
\begin{align}
    p_\text{T}^{b_1b_2},~p_\text{T}^{\gamma_1\gamma_2}>100\gev.
\end{align}
With these cuts, we have $\kappa_{\rm select}=0.90$.

\begin{figure}
	 \centering
	\includegraphics[scale=0.6]{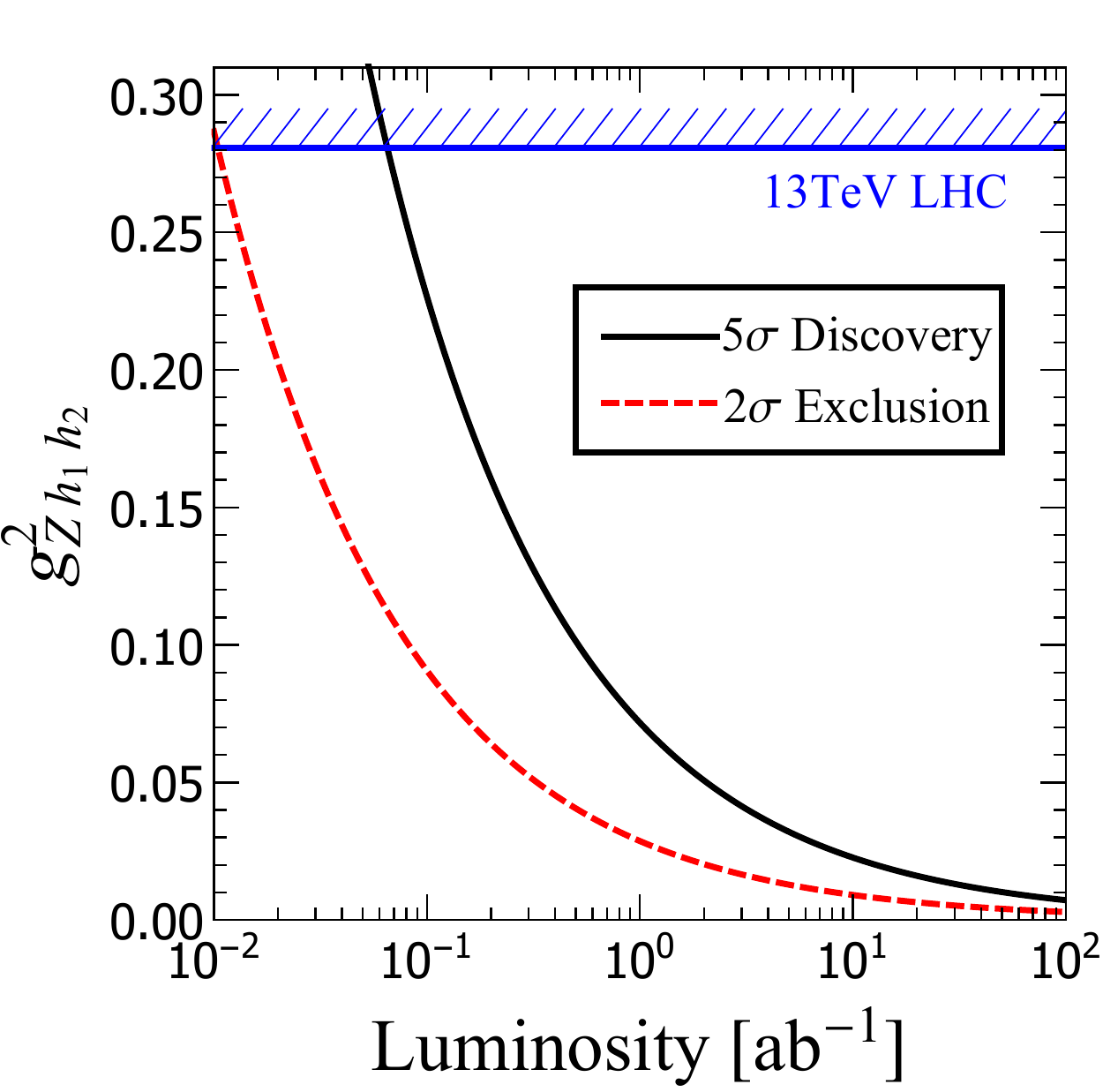}\\
    \caption{The contours of the discovery potential of $pp \to Z^*\to h_1h_2$ at the FCC-hh in the plane of $g_{Zh_1h_2}^2$ and the integrated luminosity, with the assumption of $\kappa_{b\bar b\gamma\gamma}^{h_1h_2}=1$. The region above the blue line is excluded by the LHC data.}
	\label{fig:Fcchh}
\end{figure}

With an integrated luminosity of 30 ab$^{-1}$, the number of signal events is
\begin{equation}
75544 \times g^2_{Zh_1h_2} \kappa^{h_1h_2}_{b\bar{b}\gamma\gamma},  
\end{equation} 
while the total background event number is 39179, including 12061 events from the ggF channel and 27118 events from the QCD backgrounds~\cite{Contino:2016spe}.

With an expected confidence level of $z$ and an integrated luminosity of $L$, the bound of $g_{Z h_1h_2}$ is
\begin{equation}
    g^2_{Z h_1h_2}=0.0718\times\frac{1}{\kappa^{h_1h_2}_{b\bar{b}\gamma\gamma}} \left(\frac{z}{5}\right)\sqrt{\frac{{1~}\mathrm{ab}^{-1}}{L}}.
    \label{eq:fcc_hh}
\end{equation}
Under the assumption of $\kappa^{h_1h_2}_{b\bar{b}\gamma\gamma}=1$, the significance of the measurement of $g_{Zh_1h_2}$ is displayed in Fig.~\ref{fig:Fcchh}. The result is greatly improved due to the large collision energy of the FCC-hh. For example, the effective coupling $g_{Zh_1h_2}$ of 0.114 can be discovered at $5\sigma$ C.L. with an integrated luminosity of 30 ab$^{-1}$.

\section{Future Lepton colliders}
\label{sec:cepc}

\begin{figure*}
	 \centering
	 \includegraphics[scale=0.62]{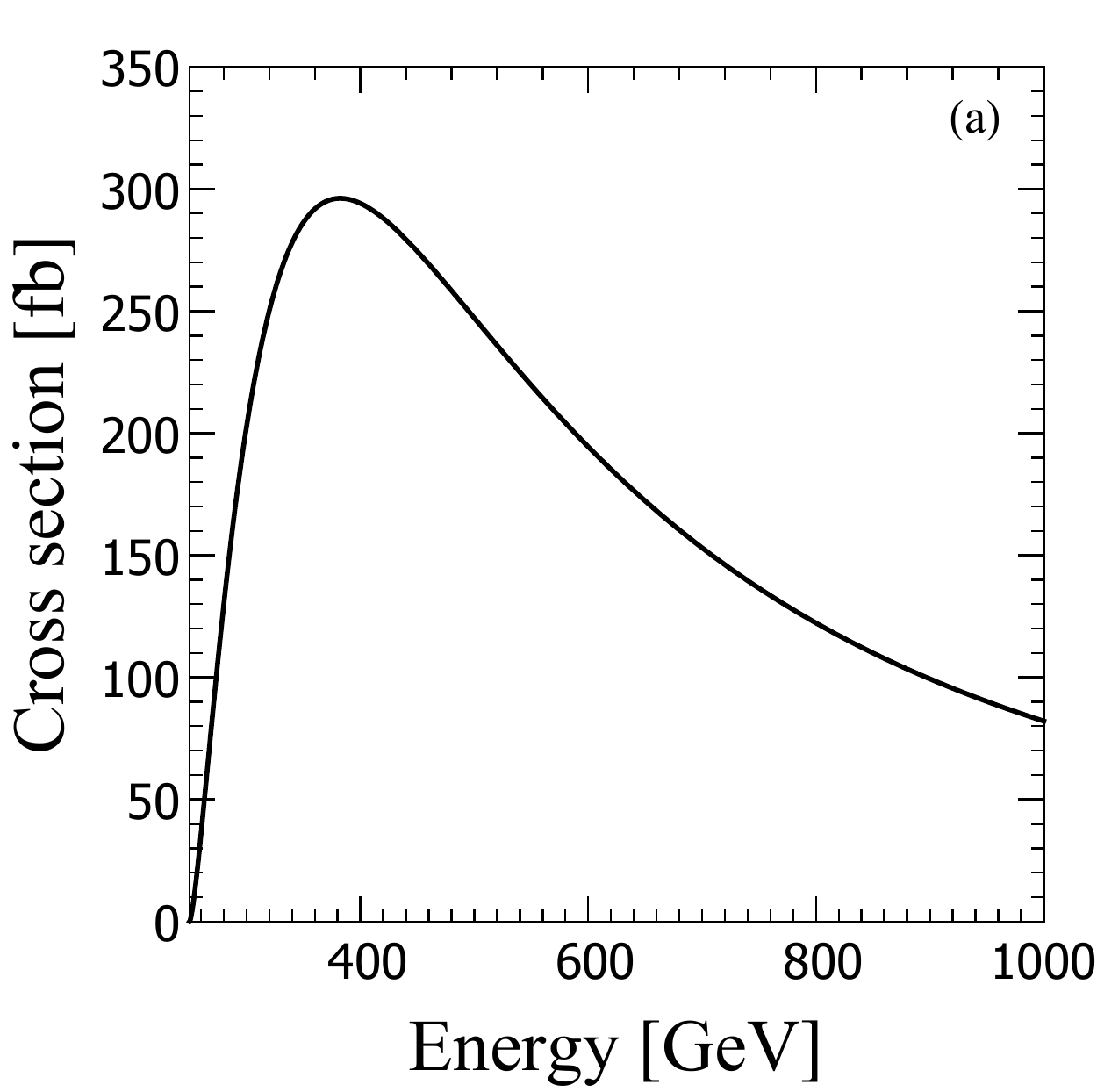}~~~~~~~~
	\includegraphics[scale=0.62]{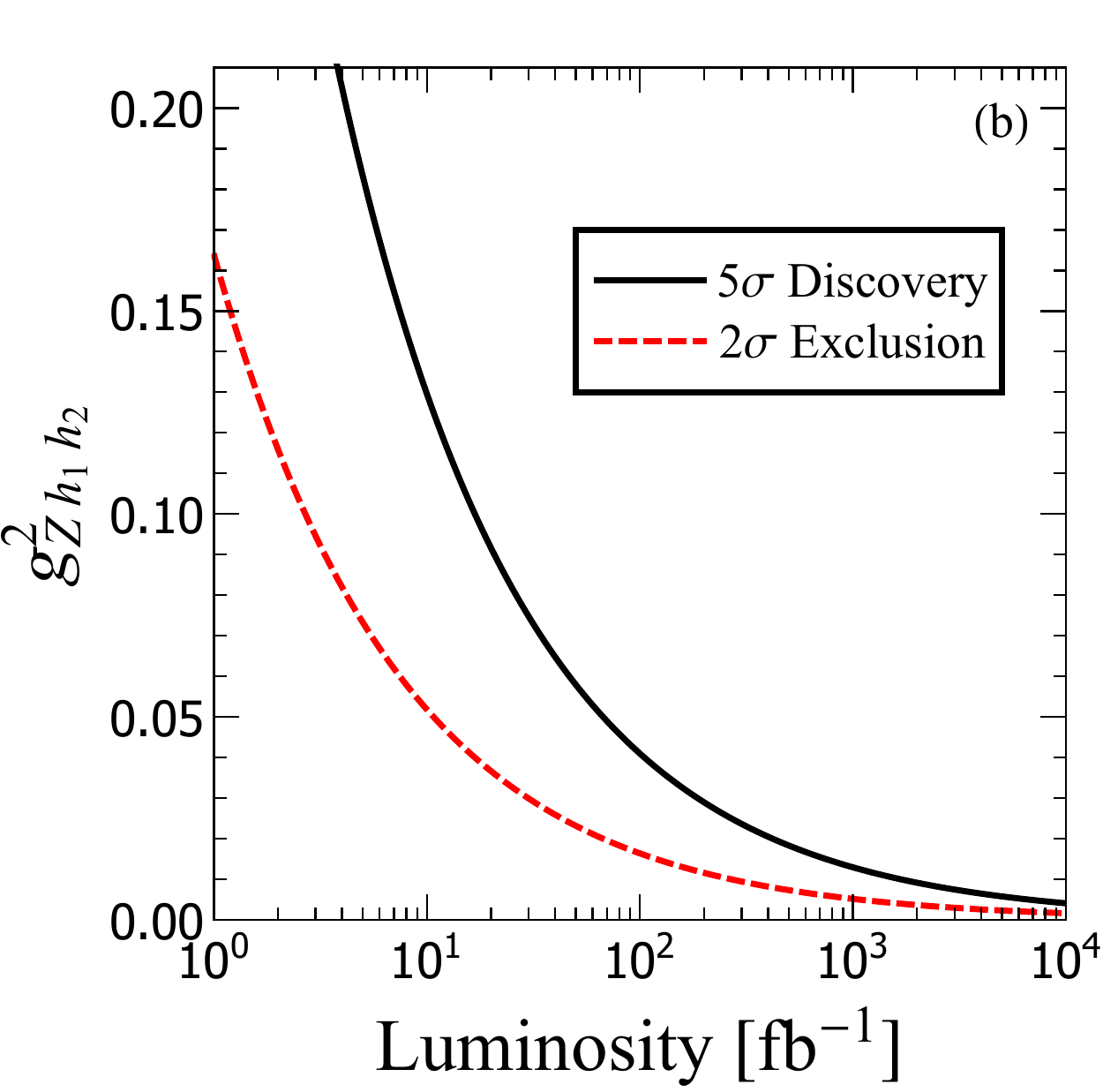}
\caption{
    (a) the cross-section of di-Higgs boson production through the process of $e^+e^-\to Z^\ast \to h_1h_2$ as a function of the collision energy dependence for $g_{Zh_1h_2}=1$; (b) the potential of a 350~GeV \eec on measuring $g_{Zh_1h_2}$ for $\kappa_{b\bar b\gamma\gamma}^{h_1h_2}=1$.}
\label{fig:CEPC}
\end{figure*}

In this section, we study phenomenology at future electron-position colliders, such as CEPC~\cite{CEPC}, FCC-ee~\cite{FCC:2018evy}, and ILC~\cite{ILC:2013jhg}.
Figure.~\ref{fig:CEPC} displays the cross-section of $e^{+}e^{-}\to Z^\ast \to h_1h_2$ as a function of the collision energy ($\sqrt{s}$) in the COM frame, which exhibits a maximum at around $\sqrt{s}=350~\rm{GeV}$. It is a result of the competition between the $s$-channel propagator suppression and the $p$-wave enhancement; see Eq.~\ref{eq:dsdcosffZhA}. We choose $\sqrt{s}=350 \gev$ as a benchmark point, and the LO production cross-section is 
\begin{align}
\sigma_{Zh_1h_2} =g_{Zh_1h_2}^2 \times 282.82 \fb.
\end{align}
Since the QCD background at lepton colliders are much less than hadron colliders, we focus on the $4b$ decay mode of $h_1h_2$.

The irreducible background arises from the loop-induced di-Higgs boson production in the SM, and its cross-section is around $0.01$ fb which is negligible. The reducible backgrounds are $e^+e^-\to Zh_1(h_2) \to b\bar{b} b\bar{b}$ and $e^+e^-\to ZZ \to b\bar{b} b\bar{b}$ with LO cross-section of 15.8 fb and 11.5 fb, respectively\footnote{The $hVV$ coupling is highly constrained by current data. We assume that the total production cross-section of $e^+e^-\to Zh_1$ and $e^+e^-\to Zh_2$ is the same as that of $e^+e^-\to Zh$ in the SM. }, calculated with $\mathtt{MadGraph5\_aMC@NLO}$~\cite{madgraph5}. The $W$-pair background is negligible for a low mistagging rate. There are also continuum QCD backgrounds $e^+e^-\to b\bar{b}b\bar{b}/b\bar{b}c\bar{c}/b\bar{b}jj$, where $j$ denotes light jets.
To avoid the soft and the collinear divergences, we require 
\begin{equation}\label{eq:precut}
  \pt^{b,j} > 20 \gev,\quad|\eta^{b,j}| < 5.0,\quad\Delta R^{bb,bj,jj} > 0.2,
\end{equation}
when we calculate the cross-section and generate the parton-level events of the QCD backgrounds.
The Durham algorithm with $d_{\mathrm {cut}}=(20\gev)^2$~$(y_{\mathrm{cut}}=0.00327)$ is chosen to reconstruct jets~\cite{Catani:1991hj,Cacciari:2011ma}. The $b$-jet (mis-)tagging efficiencies at CEPC are given as~\cite{CEPC}
\begin{equation}
    \epsilon_{b\to b}=0.80,\quad
    \epsilon_{c\to b}=0.1,\quad
    \epsilon_{j\to b}=0.01.
\end{equation}
We require the events to contain at least four jets, and the leading four jets must be $b$-tagged hard jets that satisfy
\begin{align}
    p_\text{T}^{b}>25~{\rm{GeV}},\quad |\eta^{b}|<2.5,\quad \Delta R^{{bb}}>0.5.
\end{align}
As the four bottom-quarks originating from two Higgs bosons in the signal events, we demand that there are two reconstructed Higgs bosons with reconstructed invariant mass in the range from 110 GeV to 140 GeV. 
To reconstruct two Higgs bosons, $\chi^2$ method is utilized, namely
\begin{align}
    \chi^2\equiv\sqrt{(m_{ij}-m_H)^2+(m_{kl}-m_H)^2}
\end{align}
is required to minimal, where $m_{ij(kl)}$ is the invariant mass of the $i(k)$th and the $j(l)$th $b$-jet. 
After going through the cuts above, the cross-section of the signal is $g_{Zh_1h_2}^2\kappa^{h_1h_2}_{b\bar{b}b\bar{b}}\times 8.2$ fb, and the production cross-section for the EW and QCD backgrounds are 0.42 fb and 0.03 fb, respectively. 

In Fig.~\ref{fig:CEPC}, we show the potential of discovery and exclusion of the effective coupling $g_{Zh_1h_2}$, with an assumption that the scalars decay branching ratio to $b\bar{b}$ is the same as the SM Higgs boson. For convenience, the effective coupling, which can be measured to a specific confidence level with an integrated luminosity of $L$, reads as
\begin{equation}
g^2_{Z h_1h_2}=0.0130\times\frac{1}{\kappa^{h_1h_2}_{b\bar{b}\gamma\gamma}}
\left(\frac{z}{5}\right)\sqrt{\frac{1~\text{ab}^{-1}}{L}},
\label{eq:cepc}
\end{equation}
under the same assumption of scalars decay branching ratio.
To obtain the same measurement accuracy, the integrated luminosity required by 350~GeV $e^+e^-$ colliders is about one-seventh of that of the FCC-hh; see Eqs.~\ref{eq:fcc_hh} and \ref{eq:cepc}.

\section{Conclusion}
\label{sec:conclusion}

Ten years after the Higgs boson discovery,  many properties of the Higgs boson, e.g., its mass, spin, CP property, and interaction with other SM particles, have been measured precisely. However, most of the time, people assume that there is only one Higgs particle with a mass of 125 GeV. In this work, we examine the possibility of another scalar hidden behind the known Higgs boson. The hidden scalar has a mass degenerate with or very close to the Higgs boson such that one cannot recognize it. Suppose the hidden scalar exhibits a CP property different from the known Higgs boson. We propose to utilize the $Zh_1h_2$ interaction to reveal the CP-odd component at the measured Higgs boson mass of 125 GeV as the vertex unambiguously indicates two degrees of scalar freedoms with different CP properties. The $Zh_1h_2$ interaction induces a new channel of di-Higgs boson production which can be tested at the colliders. We discuss the (HL-)LHC potential on the di-Higgs boson production, including the electroweak and gluon-gluon fusion contributions. It is found that the di-Higgs boson production can be discovered for $g_{Zh_1h_2}=0.37$ at the $5\sigma$ confidence level.

While at the future colliders, we focus on the precision measurement ability of the effective coupling $g_{Zh_1h_2}$. It is found that the effective coupling of $g_{Zh_1h_2}=0.114$ can be confirmed at the FCC-hh with an integrated luminosity of 30 ab$^{-1}$ at $5\sigma$ confidence level. However, a less integrated luminosity is needed at the \eec because of the clean environment; for example, it requires an integrated luminosity of 5 ab$^{-1}$ at the \eec to reach the same accuracy as at the FCC-hh. 

Our study is based on the effective coupling $g_{Zh_1h_2}$ rather than focusing on a specific new physics model. In a complete UV model, the hidden scalar might appear in the single Higgs boson production and be constrained by current experimental data. However, it highly depends on the Yukawa interactions of the hidden scalar to the SM fermions and other new physics particles. The detailed discussion in specific new physics models will be presented elsewhere~\cite{Cao:2022pre}.

\begin{acknowledgments}
The work is partly supported by the National Science Foundation of China under Grant Nos. 11725520, 11675002, 11635001, 11805013, 12075257, 12235001, and the funding from the Institute of High Energy Physics, Chinese Academy of Sciences (Y6515580U1) and the funding from Chinese Academy of Sciences (Y8291120K2, E12D172901).
\end{acknowledgments}

\bibliographystyle{apsrev}
\bibliography{reference}
\end{document}